\documentclass[lettersize,journal]{IEEEtran}
\usepackage{amsmath,amsfonts}
\usepackage{algorithmic}
\usepackage{algorithm}
\usepackage{array}
\usepackage{textcomp}
\usepackage{stfloats}
\usepackage{url}
\usepackage{verbatim}
\usepackage{graphicx}
\usepackage{float}
\usepackage{subfigure}
\usepackage{cite}
\usepackage{booktabs}
\usepackage{multirow}

\usepackage{hyperref}
\hypersetup{hypertex=true,
colorlinks=true,
linkcolor=blue,
anchorcolor=blue,
citecolor=blue}

\renewcommand{\algorithmicrequire}{\textbf{Input:}}
\renewcommand{\algorithmicensure}{\textbf{Output:}}

\begin{document}
\definecolor{purple}{rgb}{0.65,0,0.65}
\definecolor{black}{rgb}{0.0, 0.0, 0.0}
\newcommand{\chen}[1]{{\color{purple}#1}}

\title{GIFDL: Generated Image Fluctuation Distortion Learning for Enhancing Steganographic Security}

\author{Xiangkun Wang, Kejiang Chen, Yuang Qi, Ruiheng Liu, Weiming Zhang, Nenghai Yu

\thanks{This work was supported in part by the Natural Science Foundation of China under Grant U2336206, 62472398, U2436601 and 62402469}
\thanks{All the authors are with School of Cyber Science and Technology, University of Science
and Technology of China and Anhui Province Key Laboratory of Digital Security, Hefei 230026, China.}
\thanks{Corresponding author: Kejiang Chen
(Email:chenkj@ustc.edu.cn)}}

\markboth{Submitted to IEEE TIFS}%
{Shell \MakeLowercase{\textit{et al.}}: A Sample Article Using IEEEtran.cls for IEEE Journals}

\maketitle

\begin{abstract}
Minimum distortion steganography is currently the mainstream method for modification-based steganography. A key issue in this method is how to define steganographic distortion. With the rapid development of deep learning technology, the definition of distortion has evolved from manual design to deep learning design. Concurrently, rapid advancements in image generation have made generated images viable as cover media. However, existing distortion design methods based on machine learning do not fully leverage the advantages of generated cover media, resulting in suboptimal security performance. To address this issue, we propose GIFDL (Generated Image Fluctuation Distortion Learning), a steganographic distortion learning method based on the fluctuations in generated images. Inspired by the idea of natural steganography, we take a series of highly similar fluctuation images as the input to the steganographic distortion generator and introduce a new GAN training strategy to disguise stego images as fluctuation images. Experimental results demonstrate that GIFDL, compared with state-of-the-art GAN-based distortion learning methods, exhibits superior resistance to steganalysis, increasing the detection error rates by an average of 3.30\% across three steganalyzers.
\end{abstract}

\begin{IEEEkeywords}
Generative adversarial networks (GANs), steganography, generative model.
\end{IEEEkeywords}

\section{Introduction}
\IEEEPARstart{I}{mage} steganography is a technique for hiding information, aiming to embed secret messages within images so that they are not easily detected~\cite{imagesteganography}. Steganalysis, a countermeasure to steganography, seeks to identify the presence of hidden messages in images~\cite{Steganalysis}. Early steganography methods, known as LSB steganography~\cite{mielikainen2006lsb}, hide secret messages in the least significant bits of pixels. These methods do not consider the different risks of performing steganographic modifications in different regions of the image, making them susceptible to detection by statistical steganalysis techniques. Content-adaptive steganography is then proposed, which adjusts the location and capacity of hidden messages based on the texture complexity of the image. The most typical achievement of this domain is minimum distortion steganography~\cite{minimum}, which separates steganography into two processes: defining steganographic distortion and steganographic encoding, offering higher security than LSB steganography. Steganographic coding schemes~\cite{stc},~\cite{spc}, and~\cite{LGDM} are approximately optimal in minimizing total distortion, thus research focuses on how to better define steganographic distortion. Methods such as WOW~\cite{ref1}, UNIWARD~\cite{ref2}, HILL~\cite{ref3}, MiPOD~\cite{ref4}, UERD~\cite{ref5}, MS~\cite{chen2018defining}, and IP~\cite{hongxia} have been proposed to improve steganographic security by rationally defining distortion.

\begin{figure}[t]\centering
\includegraphics[width=\columnwidth]{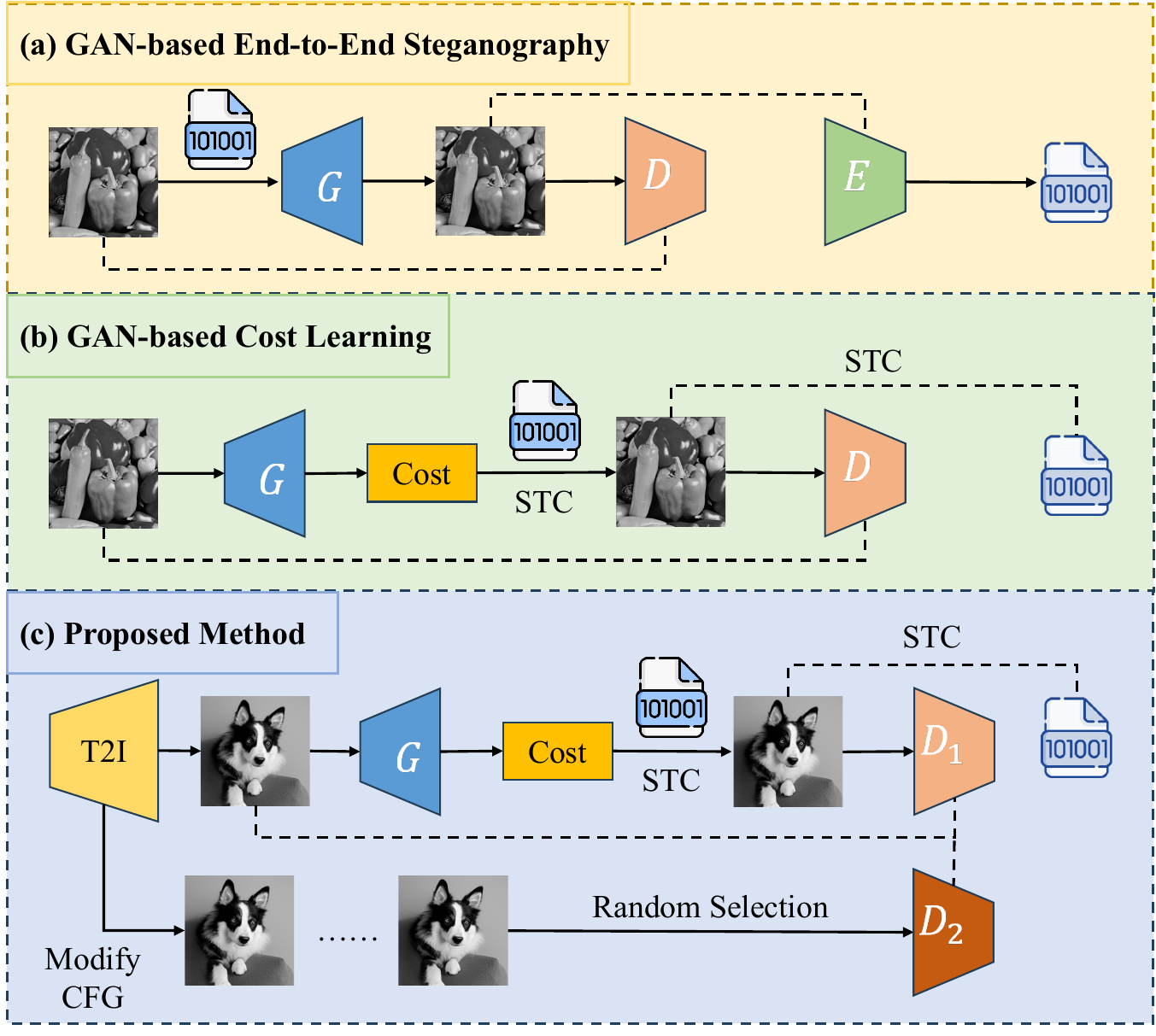}
\caption{Comparison of existing GAN-based steganography methods with the proposed method.}
\label{pintro}
\end{figure}

With the development of deep learning, traditional steganographic methods are not secure enough against deep learning-based steganalysis. Moreover, the excellent performance of generative adversarial networks (GANs) in image processing has inspired new steganography methods. GAN-based image steganography methods can be categorized into two types: GAN-based end-to-end steganography and GAN-based cost learning. As shown in Fig.~\ref{pintro}(a), GAN-based end-to-end steganography does not employ the minimum distortion steganography framework. Instead, it explores the direct generation of stego images from cover images via the generator. For instance, SteganoGAN~\cite{steganogan} embeds secret messages within the feature maps of cover images, achieving a payload of up to 4 bpp (bits per pixel). ABDH~\cite{attentionyu} introduces attention mechanisms in GANs to improve the quality and robustness of stego images. Building on SteganoGAN and ABDH, ChatGAN~\cite{chatgan} further enhances image quality using XuNet~\cite{xunet} as the discriminator. However, these end-to-end methods suffer from significant security weaknesses against steganalysis. Deep learning-based steganalyzers such as SRNet~\cite{SRnet} can accurately detect stego images via a small number of training images.

Unlike GAN-based end-to-end steganography, GAN-based cost learning methods adhere to the minimum distortion steganography framework. They use GANs to learn steganographic distortions, convert them into steganographic costs, and combine steganographic encoding to embed and extract secret messages, as shown in Fig.~\ref{pintro}(b). ASDL-GAN~\cite{ASDL-GAN} is the first to apply GANs to learn steganographic distortion, using XuNet as the discriminator, but its security is weaker than that of manually designed distortion methods~\cite{ref3}. UT-GAN~\cite{ut-gan} improved upon ASDL-GAN by adopting a U-Net architecture in the generator and adding a double-tanh function, achieving better security than manual distortion definitions. GMAN~\cite{GMAN} highlights that the security of GAN-based steganography is limited by discriminator performance, and a multi-discriminator training strategy is proposed to enhance security further. GACL~\cite{wang2024gan} enhances the cover image to highlight its contour information, which in turn improves steganographic security. Despite these advancements, GAN-based distortion learning methods still face security challenges at high payloads, and designing more secure steganographic methods remains a crucial objective for researchers.

The development of deep learning has provided us with not only better cost learning methods but also new types of cover media. According to the Everypixel Journal\footnote{\!\url{https://journal.everypixel.com/ai-image-statistics}}, text-to-image generation models have created over 15 billion images during just a year and a half, equivalent to the number of photos taken by humans over 149 years. The widespread popularity of generated images makes them ideal steganographic covers. Researchers have proposed several steganographic methods based on generated images. For example, IDEAS~\cite{ideas} synthesizes stego images by generating image structural features from secret messages. GSN~\cite{gsn} uses secret messages as latent variables to guide the generator in image synthesis. StegaDDPM~\cite{stegaddpm} generates stego images by sampling noise according to secret messages in the final step of the denoising diffusion probabilistic model. These methods require specially designed generator architectures or access to the internal parameters of the generation model to control the image generation process. However, popular image generation models are often considered black-box, with users having no access to internal information. Recently, Zhang et al.~\cite{zjs} proposed volatility distortion (VC) based on the fluctuation of black-box generative models, which can be used to improve the security of existing steganographic distortion learning methods by approximating the image pixel distribution using a manual method. However, the hand-designed method may make it difficult to comprehensively and accurately model the fluctuation distribution of an image, and further improvements are possible.

To address this issue, we propose a new GAN-based steganographic cost learning method, named GIFDL (Generated Image Fluctuation Distortion Learning), as shown in Fig.~\ref{pintro}(c). We observe that when the input parameters of an image generation model are slightly modified, it outputs images that are highly similar to the original, which we call fluctuation images. Inspired by the idea of natural steganography~\cite{naturalsteganography}, the goal of GIFDL is to disguise stego images as fluctuation images. To this end, we take a series of fluctuation images as inputs to the generator, select one of them as the cover image, and randomly select one of the remaining fluctuation images, called ``flu''. We pursue the stego image indistinguishable from fluctuation images, so GIFDL employs two discriminators, one for distinguishing between stego and cover, and the other for distinguishing between stego and ``flu'', with the two discriminators updating their parameters alternately. Considering that the cover is also a member of the fluctuation images, the joint function of the two discriminators can ensure that the stego image is close to fluctuation images in terms of pixel distribution.

With the above design, the stego images resemble the fluctuation images, enabling the generator to learn the fluctuation distribution of the generated images, which allows the generator to modify pixels that deviate from the distribution at a higher cost. Under the minimum distortion steganography framework, we ultimately obtain stego images that are close to the fluctuation images in distribution. Unlike other GAN-based steganography methods, GIFDL focuses on the characteristics of black-box generated images and leverages their fluctuations to achieve steganographic distortion learning. Our contributions can be summarized as follows:

\begin{itemize}

\item[$\bullet$] \textbf{A novel steganographic distortion learning method based on the fluctuations of generated images}: Based on the observation of the fluctuation characteristics of the generated images, we take the fluctuation images as as the input of our network and use a GAN to learn the distribution of fluctuation images, which in turn disguises the stego images as fluctuation images.

\item[$\bullet$] \textbf{A new discriminator training strategy that considers the inherent differences between cover and fluctuation images}: We use a steganalysis network to distinguish between cover and stego images, while another steganalysis network is used to distinguish between fluctuation and stego images. By updating these two discriminators alternately, GIFDL can better learn the characteristics of fluctuation images, achieving better camouflage.

\item[$\bullet$] \textbf{Extensive experiments to validate the effectiveness of GIFDL}: Extensive experiments have shown that GIFDL achieves considerable security against three mainstream steganalyzers. Compared with state-of-the-art methods, the detection error rate of steganalysis is improved by 3.30\% on average.

\end{itemize}

\begin{figure}[t]\centering
\includegraphics[width=\columnwidth]{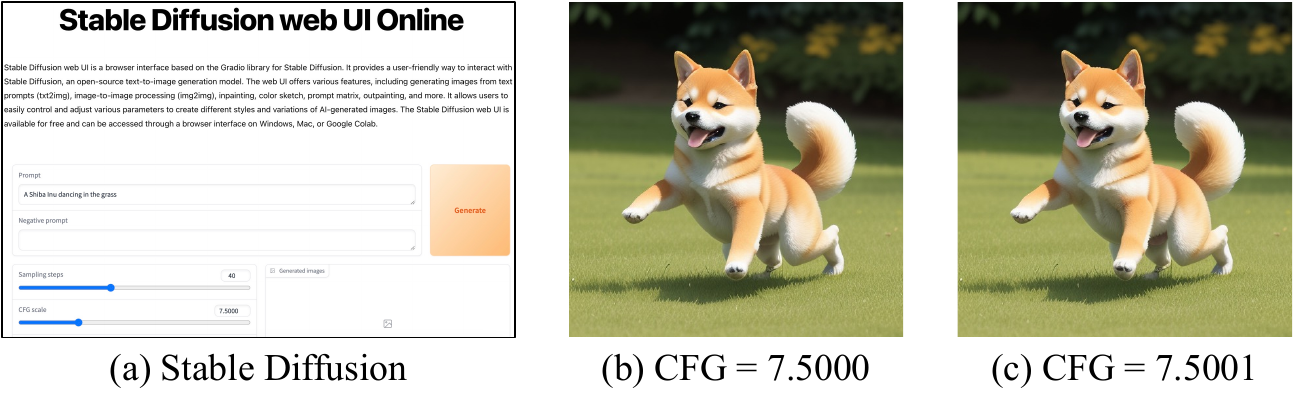}
\caption{(a) Users can access the black-box Stable Diffusion at \href{https://stabledifffusion.com/webui}{Stable Diffusion web UI Online} to use Stable Diffusion. (b) Images generated with CFG = 7.5000. (c) Images generated with CFG = 7.5001, where the Classifier-Free Guidance (CFG) scale is an input parameter in Stable Diffusion that controls how closely a prompt should be followed.}
\label{psd}
\end{figure}

\section{Related Work}
\label{c1}
In this section, we first introduce GAN-based cost learning methods, then we introduce volatility cost, followed by an introduction to text-to-image generation models.
\subsection{GAN-based cost learning}
The GAN consists of two adversarial sub-networks: the generator, which aims to create samples that resemble real data, and the discriminator, which aims to distinguish between real data and fake samples generated by the generator. GAN-based cost learning methods are built on the minimum distortion steganography framework, which uses a GAN to generate steganographic costs and employs steganographic encoding to embed and extract secret messages. GAN-based cost learning methods leverage the adversarial nature of GANs and use deep-learning steganalyzers as discriminators. During training, these methods learn the implicit features of steganalysis algorithms within the discriminator and feed this information back to the generator to produce a modification probability map, thus obtaining steganographic cost. Notable methods include ASDL-GAN~\cite{ASDL-GAN}, UT-GAN~\cite{ut-gan}, GMAN~\cite{GMAN} and GACL~\cite{wang2024gan}.

\subsubsection{ASDL-GAN}
ASDL-GAN is the first to propose using the adversarial nature of GANs to learn steganographic cost, employing the steganalysis network Xu-Net as the discriminator. The generator learns the modification probabilities of the cover image, which are then mapped to modifications through a ternary embedding simulator (TES). The loss function of the generator is directly related to the undetectability by the adversarial steganalyzer and achieved backpropagation through TES. However, the security performance was not satisfactory. Experimental results revealed that ASDL-GAN was less effective against steganalysis compared with manually designed methods such as HILL.

\subsubsection{UT-GAN}
UT-GAN builds upon ASDL-GAN by employing a generator based on the U-Net architecture. Additionally, it introduces a dual-tanh activation function that does not require training, replacing the TES in ASDL-GAN. This improvement means that steganographic performance is no longer constrained by the pre-training of TES, and that the training time of GAN is reduced. Experimental results demonstrate that UT-GAN outperforms ASDL-GAN and the manual method HILL in resisting steganalysis.

\subsubsection{GMAN}
GMAN addresses the limitations of ASDL-GAN and UT-GAN, which use a single, relatively weak steganalyzer, Xu-Net, as the GAN discriminator. A weak discriminator can limit the generator's performance, thereby constraining the overall security of the steganographic method. GMAN introduces multiple steganalyzers as discriminators. During training, GMAN adaptively updates the parameters of the discriminators based on their performance. Experimental results indicate that GMAN achieves the highest security performance among the methods tested.

\subsubsection{GACL}
GACL enhances the cover image using Laplace operators to highlight the edges and contours of the image to obtain an enhanced image. GACL then learns the steganographic distortion by feeding the cover image together with the enhanced image into a two-stream U-Net network.

These methods focus on optimizing the design of the network structure of the GAN itself and pay less attention to steganographic cover.

\subsection{Volatility Cost}
On the basis of the fluctuation of generated images, Zhang et al.~\cite{zjs} first proposed the idea of disguising the steganographic modification as the inherent volatility of the generated model, which improves the steganographic security and can be used in the black-box scenarios of the generated model. They approximated the pixel distribution of fluctuation images as a Gaussian distribution and estimated the parameters of the Gaussian distribution through a series of fluctuation images, which were then integrated to obtain the occurrence probability of each pixel. Furthermore, Volatility Cost (VC) was introduced based on the estimated distribution, with the probability of generated pixel occurrences as the steganographic modification probability, thus translating into volatility cost. Since VC calculates distortion only in terms of image volatility, for comprehensive consideration, Since VC calculates distortion solely in terms of image volatility, for a more comprehensive approach, it was combined with existing distortion definitions, improving security by an average of 4.64\%. However, the distortion definition method of VC is based on some approximation assumptions, which inevitably have a bias in estimating the pixel distribution of fluctuation images and cannot comprehensively model the fluctuation characteristics of the generated images. Considering the limitations of VC, we propose to use deep learning methods to learn the fluctuation characteristics of generated images.

\subsection{Text-to-image Generative Model}
Text-to-image (T2I) generation models take text descriptions as inputs and produce high-quality, realistic images that correspond to the given descriptions. To increase image quality, numerous methods have been developed. For instance, Text-conditional GAN~\cite{text_condition} was the first to implement an end-to-end differentiable architecture from the character level to the pixel level. StackGAN~\cite{stackgan} utilizes multiple stacked generators, AttnGAN~\cite{attngan} incorporates attention mechanisms, and ControGAN~\cite{controlgan} employs a word-level discriminator and perceptual loss to control image generation. Recently, Denoising Diffusion Probability Models (DDPMs) have emerged as a new paradigm in image generation due to their outstanding performance. DDPM-based models such as Imagen~\cite{imagen}, Stable Diffusion~\cite{stablediffusion}, and DALL-E 3~\cite{dalle3} can generate images that are close to real images while also having artistic qualities. According to Everypixel Journal, in 2024, people created an average of 34 million images per day, which are widely shared on social networks.

\section{Proposed Method}
\label{c2}
\subsection{Motivation}
\label{c2-1}

\begin{figure}[t]\centering
\includegraphics[width=\columnwidth]{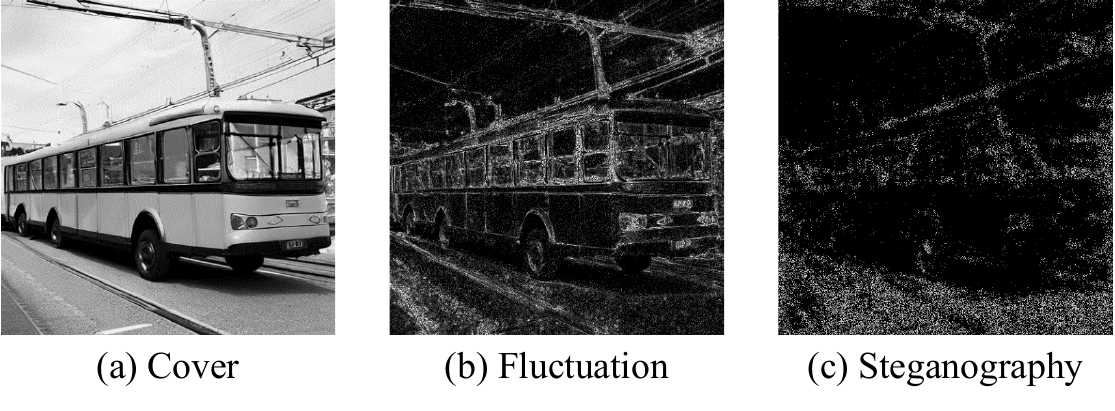}
\caption{(a) Cover image, (b) average pixel difference between the 10 fluctuation images and the cover image, (c) pixel difference between the stego image and the cover image, where the stego image is generated by GMAN. For clearer observation, the brightness of (b) and (c) is multiplied by 50.}
\label{pmoti}
\end{figure}

\begin{figure*}[t]
\centerline{\includegraphics[width=\textwidth]{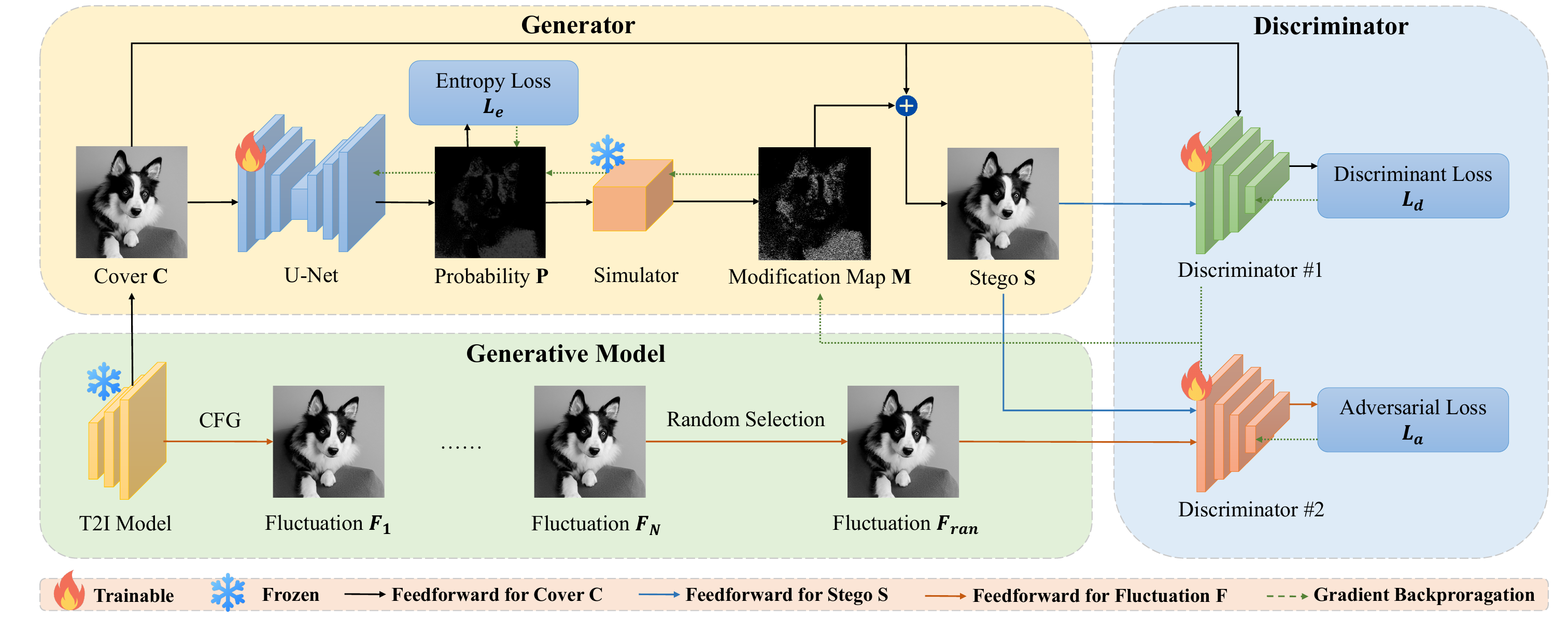}}
\caption{The overall framework of the proposed GIFDL consists of three components: the Generative Model, the Generator, and the Discriminator. In the Generative Model, we use a T2I model to generate cover image $C$ and corresponding fluctuation images $F_i, i = 1, 2, \ldots$ From this series of fluctuation images, a random fluctuation image $F_{ran}$ is selected to participate in the subsequent training of the generator and discriminator. In the Generator, we employ a U-Net architecture to generate the modification probability map $P$. Using a simulation embedder, we obtain the modification map $M$, which is further used to produce the stego image $S$. In the Discriminator, Discriminator \#1 is used to distinguish between the cover image $C$ and the stego image $S$, while Discriminator \#2 distinguishes between the fluctuation image $F_{ran}$ and the stego image $S$. It is important to note that $F_{ran}$ is randomly selected in each training epoch and is not fixed.}
\label{frame}
\end{figure*}

Research in the field of image steganography has consistently shown that introducing steganographic modifications in areas with complex textures is less likely to be detected. Based on this insight, researchers have proposed heuristic steganographic distortion methods such as WOW~\cite{ref1}, S-UNIWARD~\cite{ref2}, HILL~\cite{ref3}, MiPOD~\cite{ref4}, and UERD~\cite{ref5}. With the advancement of deep learning, researchers have also developed GAN-based methods to learn steganographic distortion definitions, including ASDL-GAN~\cite{ASDL-GAN}, UT-GAN~\cite{ut-gan}, GMAN~\cite{GMAN} and GACL~\cite{wang2024gan}. In these methods, the GAN's discriminator is composed of steganalysis networks, and the generator learns the definition of steganographic distortion based on the adversarial loss from the discriminator. This means that the learning of steganographic distortion is driven solely by the adversarial interaction with the steganalysis network.

However, as demonstrated in GMAN, current methods typically use the relatively weak steganalysis network XuNet as the discriminator. Stronger steganalysis networks such as Yed-Net can lead to gradient vanishing and training failure. These factors limit the performance of the generator. By introducing a new discriminator training strategy, GMAN employs multiple steganalyzers as discriminators to increase the security of stego images. Nonetheless, experimental results revealed that using three steganalyzers was less effective than using two, indicating that steganographic security does not necessarily improve with an increasing number of steganalyzers and can even decrease. Therefore, we pose the following question: In scenarios where improving the discriminator's performance does not enhance steganographic security, how can GAN-based distortion learning methods further improve security?

We have observed that generated images exhibit certain fluctuation characteristics. For example, in Stable Diffusion, the input parameter ``Classifier-Free Guidance"(CFG) scale controls the similarity between the text and the image. This value is typically set to 7 to balance similarity and image quality. When the CFG value is slightly changed, Stable Diffusion outputs two nearly identical images. As shown in Fig.~\ref{psd} (b) and (c), when the CFG changes from 7.5000 to 7.5001, the differences between the two generated images are imperceptible to the human eye. To further explore the distribution properties of fluctuation images, we used Stable Diffusion to generate 11 images with CFG values of 7.5000, 7.5010, 7.5020,..., and so on. We selected one of these images as the cover image and the remaining 10 as fluctuation images, using GMAN to generate the stego image corresponding to the cover image. Fig.~\ref{pmoti} (b) shows the average pixel differences between the cover image and the 10 fluctuation images, and Fig.~\ref{pmoti} (c) shows the pixel difference between the cover and the stego image. For better visualization, the brightness in Fig.~\ref{pmoti} (b) and (c) has been increased by 50 times. As depicted in Fig.~\ref{pmoti}, the pixel differences between fluctuation images and the cover image are concentrated in areas with complex textures, which closely aligns with the distribution of modifications introduced by steganography.

Based on these observations, we propose that the fluctuation characteristics of generated images can provide distribution information about the images, which we term the fluctuation distribution. \textbf{Inspired by the idea of natural steganography~\cite{naturalsteganography}, we aim to disguise stego images as fluctuation images, i.e., modifications of steganography, disguised as fluctuation differences.} Therefore, we propose a steganographic cost learning method based on the fluctuations of generated images, named GIFDL. By using fluctuation images to provide side information, we ensure that stego images are not only indistinguishable from cover images but also difficult to distinguish from fluctuation images, which makes the distribution of stego images similar to that of fluctuation images, disguising the stego images as those produced under slight input parameter fluctuations.

\subsection{Framework}
\label{c2-b}

As shown in Fig.~\ref{frame}, the overall framework of the proposed GIFDL consists of three components. In GIFDL, the U-Net of the Generator, as well as the two discriminators are trainable, while the T2I Model and the Simulator are fixed.

\subsubsection{Generative Model}

\begin{figure}[t]\centering
\includegraphics[width=\columnwidth]{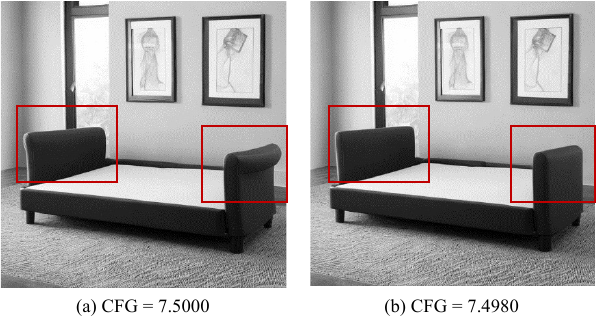}
\caption{(a) Image generated with CFG = 7.5000. (b) Image generated with CFG = 7.4980. There are significant differences between (a) and (b).}
\label{pgm}
\end{figure}

The generative model includes a T2I Model, for which we use the pre-trained Stable Diffusion as the T2I Model to generate the cover image $C$ from the input text. We represent $C$ as $C = (c_{i,j})^{H\times W} (1\leq i\leq H, 1\leq j\leq W)$, a grayscale image with height $H$ and width $W$. By slightly altering the input parameter ``CFG scale'', we obtain a series of fluctuation images $F_1, F_2, \ldots F_N$, where $F_k = (F^k_{i,j})^{H\times W}$, $k = 1, 2, \ldots, N$. In this paper, we set $N = 10$.

The T2I model is used solely for generating cover images and the corresponding fluctuation images; it does not participate in the training process. During training, the cover image is input into the generator to produce the stego image. In each batch, one image is randomly selected from the fluctuation images, denoted as $F_{ran} = (F^{ran}_{i,j})^{H\times W}$. $F_{ran}$ is input into the discriminator to provide side information on the fluctuation distribution of the generated images.

We observed that fluctuation images $F_k$ sometimes show significant differences from the cover image $C$, as shown in Fig.~\ref{pgm}. When the difference between the cover image and the fluctuation image is too large, constraining the stego image to be similar to both at the same time becomes impossible. These large differences may provide incorrect information to the generator. Therefore, we introduce a parameter $\tau$ to limit the distance between the cover image and the fluctuation images:
\begin{equation}
MSE(C, F_k) \leq \tau,
\end{equation}where for two images $X$ and $Y$, the Mean Squared Error (MSE) is defined as follows:
\begin{equation}
MSE = \frac{1}{N} \sum_i \sum_j (X(i,j) - Y(i,j))^2,
\end{equation}where $N$ is the total number of pixels in the image, and $X(i,j)$ and $Y(i,j)$ are the pixel values of images $X$ and $Y$ at position $(i, j)$, respectively. A smaller MSE indicates that the two images are more similar, meaning that their differences are smaller.

During training, if $F_{k}$ and $C$ satisfy the above MSE condition, the corresponding sample will participate in the subsequent training process; otherwise, $F_{k}$ will be regenerated with a different ``CFG'' scale.

\subsubsection{Generator} The generator takes the cover image as input and outputs the stego image, which can be divided into two parts:

\textbf{(1) Generate the modification probability map}: This process converts the input cover image $C = (c_{i,j})^{H \times W}$ into an embedding probability $P = (p_{i,j})^{H \times W}$, where $p_{i,j}$ represents the probability that the pixel value $c_{i,j}$ will be modified by $\pm 1$ due to message embedding at pixel location $(i, j)$. In this work, we constrain the probability of each pixel value $c_{i,j}$ being modified by +1 (i.e., $p_{i,j}^{+1}$) to be equal to the probability of being modified by -1 (i.e., $p_{i,j}^{-1}$), as shown below:

\begin{equation}
p_{i,j}^{+1} = p_{i,j}^{-1} = \frac{p_{i,j}}{2}.
\end{equation}

Following the setup in UT-GAN~\cite{ut-gan}, we utilize a U-Net-based architecture for this image-to-image task (i.e., from $C$ to $P$). The U-Net consists of 15 blocks and a deconvolution layer. The first 8 blocks perform down-sampling, and the remaining 7 perform up-sampling. The output of the $i$-th block ($i = 1, 2, \ldots, 7$) is concatenated with the output of the $(16 - i)$-th block and passed into the $(17 - i)$-th block. Each block contains two convolutional layers, followed by batch normalization and LeakyReLU activation.

\textbf{(2) Simulated embedding}: This step simulates STC embedding to generate the stego image based on the probabilities $P$ obtained in the first step. First, random noise $R = (r_{i,j})^{H \times W}$ is generated, where $r_{i,j}$ is independently and identically distributed (i.i.d.) over a uniform interval $[0, 1]$. Comparing $p_{i,j}$ and $r_{i,j}$, the modification map $M = (m_{i,j})^{H \times W}$ in the simulated embedding is defined as follows:
\begin{equation}
m_{i,j} = \begin{cases}
   -1, & \text{if } r_{i,j} < p_{i,j}^{-1}, \\
   +1, & \text{if } r_{i,j} > 1 - p_{i,j}^{+1}, \\
   0, & \text{otherwise}.
   \end{cases}
\end{equation}

Since the above piecewise function is non-differentiable, we follow the setup in UT-GAN~\cite{ut-gan} and use the Double-Tanh function as an embedding simulator during training to obtain the modification map $M = (m_{i,j})^{H \times W}$. The double-tanh function serves as a differentiable approximation of the discrete piecewise function, which facilitates the backpropagation process in the GAN. The double-tanh function is defined as:
\begin{equation}
\begin{split}
m_{i,j} &= \frac{1}{2} \tanh(\gamma (p_{i,j}^{+1} - r_{i,j})) \\
&- \frac{1}{2} \tanh(\gamma (p_{i,j}^{-1} - (1 - r_{i,j}))),
\end{split}
\end{equation}where, $m_{i,j}$ in the interval $[-1, 1]$ represents the simulated modification amount, and the parameter $\gamma$ controls the precision of the simulation. We set $\gamma = 60$ as in UT-GAN. Finally, we obtain the stego image as $S = C + M$.

Once the training is complete, the embedding costs $\rho$ for the input image $C$ can be calculated as follows:
\begin{equation}
\begin{split}
\begin{cases}
&\rho_{i,j}^{+1} = \ln\left(\frac{1}{p_{i,j}^{+1}} - 2\right), \\
&\rho_{i,j}^{-1} = \ln\left(\frac{1}{p_{i,j}^{-1}} - 2\right), \\
&\rho_{i,j}^{0} = 0. 
\end{cases}
\end{split}
\end{equation}

Based on the above embedding costs, we then use STC, rather than the embedding simulator, to perform the actual message embedding, resulting in the final stego image $S$ while minimizing the total embedding cost.

\subsubsection{Discriminator}
\label{c2-c}

The discriminator consists of two distinct steganalysis networks, namely Discriminator \#1 and Discriminator \#2. Discriminator \#1 differentiates between cover images and stego images, while Discriminator \#2 distinguishes between fluctuation images $F_{ran}$ and stego images. Experimental observations indicate that steganalysis networks generally find it easier to distinguish between fluctuation images $F_{ran}$ and stego images than between cover images and stego images. Consequently, we empirically select the more powerful steganalysis network, Yed-Net~\cite{yednet}, as Discriminator \#1, and the less powerful Xu-Net~\cite{xunet} as Discriminator \#2. We name this strategy of using two different discriminators to perform two different categorization tasks as ``assignment". We will explore the design rationale of ``assignment" in Section~\ref{aban} and Section~\ref{d2}.

To balance the performance of the two discriminators given their different classification tasks, we apply the training strategy from GMAN~\cite{GMAN}, which updates the parameters of the weaker discriminator in each iteration, preventing the vanishing gradient problem during training. The specific training process of the discriminators can be divided into the following two steps:

\textbf{(1) Compute cross-entropy}: For each steganalysis network $D_i$ within the discriminators, the output of $D_i$ is the softmax layer's classification result, represented as a two-dimensional normalized vector. The classification performance of the steganalysis network $D_i$ is evaluated by binary cross-entropy $E_i$, which is defined as follows:
\begin{equation}
\label{eq7}
E_1(C,S) = -z_0 \log(D_1(C)) - z_1 \log(D_1(S)),
\end{equation}
\begin{equation}
\label{eq8}
E_2(F_{ran},S) = -z_0 \log(D_2(F_{ran})) - z_1 \log(D_2(S)),
\end{equation}where, $D_1(C), D_1(S)$ denote the classification scores of discriminator \#1 for cover $C$ and stego $S$ respectively, and $D_2(S), D_2(F_{ran})$ denote the classification scores of discriminator \#2 for stego $S$ and fluctuation $F_{ran}$ respectively. The vectors $z_0 = (1, 0)^T$ and $z_1 = (0, 1)^T$ are the true labels.

\textbf{(2) Update Discriminators and Generator}: A larger binary cross-entropy indicates a weaker steganalysis network. Suppose in a particular iteration that $E_2 \leq E_1$, which suggests that Discriminator \#1 is weaker than Discriminator \#2. Note that the classification performance of the steganalysis networks may vary across iterations. To enhance discriminative performance, we only update the parameters of the weaker steganalysis network (i.e., Discriminator \#1) in each iteration, keeping the stronger network (i.e., Discriminator \#2) unchanged. This method enables the steganalysis networks to improve gradually and progressively. To further enhance the generator's performance, both discriminators are used to guide its updates in each iteration.

\begin{algorithm}[t]
    \renewcommand{\algorithmicrequire}{\textbf{Input:}}
    \renewcommand{\algorithmicensure}{\textbf{Output:}}
	\caption{Training Steps of GIFDL}
	 \label{algorithm1}
	\begin{algorithmic}[1]
	\REQUIRE cover image $C$, fluctuation image $F_i$, $i\in \{1, ..., N\}$, number of epochs $N_e$, learning rate $\eta$.   
    \ENSURE  $\theta_g$, $\theta_{d_1}$, $\theta_{d_2}$.   
    \STATE Initialize the generator $G$, and the discriminator$D_1, \ D_2$
    \FOR {$epoch$ in $\{1, ..., N_e\}$}
    
    \STATE $P \gets  G(C;\theta_g)$
    \STATE Compute $M$ by double-tanh function
    \STATE $S \gets C + M$
    \STATE Randomly select $F_{ran} \in \{F_1, ..., F_N\}$
    \STATE Compute $E_1(C,S)$ and $E_2(F_{ran},S)$ by Eq.~\ref{eq7} and Eq.~\ref{eq8}.
    \IF {$E_1\leq E_2$} 
    \STATE Update $D_2$ by $\theta_{d_2} \gets \theta_{d_2} - \eta \nabla_{\theta_{d_2}} E_2$
    \ELSE
    \STATE Update $D_1$ by $\theta_{d_1} \gets \theta_{d_1} - \eta \nabla_{\theta_{d_1}} E_1$
    \ENDIF
    \STATE Compute $l_a = E_1 + \lambda E_2$
    \STATE Compute $l_e$ by Eq.~\ref{eq11}
    \STATE $l_G \gets -\alpha l_a + \beta l_e$
    \STATE Update $G$ by $\theta_g \gets \theta_g - \eta \nabla_{\theta_{g}} l_G$
    \STATE Learning rate $\eta$ decay
    \ENDFOR
   \STATE \textbf{return} $\theta_g$, $\theta_{d_1}$, $\theta_{d_2}$.
   \end{algorithmic} 
\end{algorithm}

\subsection{Loss Function}
\label{loss}
Below are the detailed descriptions of the loss functions in GIFDL, including the discriminator loss $l_{D_1}$, $l_{D_2}$ and the generator loss $l_G$:

\textbf{(1) Discriminator loss $l_{D_1}$ and $l_{D_2}$:}
The loss functions of Discriminator \#1 and Discriminator \#2 are defined as $l_{D_1}$ and $l_{D_2}$ respectively:
\begin{equation}
   l_{D_1}  =-z_0 \log(D_1(C)) - z_1 \log(D_1(S)),
\end{equation}
\begin{equation}
   l_{D_2}  =-z_0 \log(D_2(F_{ran})) - z_1 \log(D_2(S)).
\end{equation}

As described in Section~\ref{c2-c}, in each iteration, only the parameters of the relatively weakly performing discriminator are updated. Thus, in each iteration, if $l_{D_1}>l_{D_2}$, the parameters of Discriminator \#1 will be updated using the loss function $l_{D_1}$, and the parameters of Discriminator \#2 will be kept unchanged. Otherwise, the parameters of Discriminator \#2 will be updated using the loss function $l_{D_2}$, and the parameters of Discriminator \#1 will be kept unchanged.

\textbf{(2) Generator loss $ l_G $:}
The generator's goal is to embed messages into cover images imperceptibly, making the resulting stego images difficult for the discriminator's steganalysis networks to detect. Accordingly, the generator loss $ l_G $ is defined as:
\begin{equation}
l_G = -\alpha \cdot l_a + \beta \cdot l_e,
\end{equation} where $l_G$ consists of two components. The first term $l_a$ represents the adversarial loss against the discriminator, and the second term, $l_e$, is the entropy loss, which ensures effective payload embedding. The weights of these terms, $ \alpha $ and $ \beta $, are set as in UT-GAN~\cite{ut-gan}, with $ \alpha = 1 $ and $ \beta = 10^{-7} $. Specifically, $ l_a $ and $ l_e $ are defined as follows:
\begin{equation}
\begin{split}
   l_a = & E_1(C,S) + \lambda E_2(F_{ran},S)\\
   = &(-z_0 \log(D_1(C)) - z_1 \log(D_1(S))) + \\
   &\lambda \cdot (-z_0 \log(D_2(F_{ran})) - z_1 \log(D_2(S))),
\end{split}
\end{equation}
\begin{equation}
\label{eq11}
l_e = -\left(\sum_{\forall i,j} \sum_{\forall m} \log_2(p_{i,j}^{(m)}) - (H \times W \times q)\right)^2,
\end{equation}where, $ H $ and $ W $ represent the height and width of the input image, respectively, with $ 1 \leq i \leq H $, $ 1 \leq j \leq W $. The variable $m$ denotes the embedding modification, where $ m \in \{-1, 0, +1\} $, and $ q $ represents the embedding payload. The parameter $ \lambda $ is used to weigh the adversarial losses from two discriminators.

During the training phase, these loss functions $ l_D $ and $ l_G $ are utilized to compute gradients and update the model parameters for both the discriminator and generator in the proposed GIFDL model. The training steps of the model are outlined in Algorithm~\ref{algorithm1}.

\subsection{The Applicable Scope of GIFDL}
The design of GIFDL depends on the fluctuation characteristics of generated images. To the best of our knowledge, we have observed that text-to-image diffusion models exhibit fluctuation characteristics in the generated images when the value of the input parameter ``CFG scale" changes. The CFG scale is used in mainstream text-to-image diffusion models, such as Stable Diffusion and Midjourney. Therefore, the applicable scope of GIFDL can be defined as text-to-image diffusion models with the CFG scale.

\section{Experiments}
\label{c3}

\subsection{Experimental Setup}
\label{c3a}

Since GIFDL uses generated images as steganographic covers, we employ the widely popular image generation model Stable Diffusion v1.4 to test GIFDL's effectiveness. Stable Diffusion can generate images that match the input text (called the ``prompt"), where the ``CFG scale" parameter controls the degree of alignment between the generated image and the input text, and the ``seed" parameter determines the specific content of the generated image. We use 1,000 categories from the ImageNet~\cite{imagenet} as prompts for Stable Diffusion, assigning each category 10 different random seeds. Additionally, the ``CFG scale" parameter is set to 7.5000. As a result, each category contains 10 images with varying content, and the resulting dataset consists of 10,000 images, each of size $512\times 512$, saved in PGM format. We name this dataset $\emph{IN}_{train}$.

In section~\ref{c2-1}, we observe fluctuations in the generation model, where Stable Diffusion, given the same prompt and seed, could produce two highly similar images by slightly adjusting the ``CFG scale" value. The pixel differences between the two images are mainly concentrated in the texture areas. Therefore, we keep the prompt and seed constant and only alter the ``CFG scale" of $\emph{IN}_{train}$, with values set to 7.4950, 7.4960, 7.4970, 7.4980, 7.4990, 7.5010, 7.5020, 7.5030, 7.5040, and 7.5050. This results in 10 fluctuation datasets, consistent in content with $\emph{IN}_{train}$, which we refer to as $\emph{IN}_{flu}$. In subsequent experiments, we train the baseline model using the $\emph{IN}_{train}$ dataset and train the proposed GIFDL using both $\emph{IN}_{train}$ and $\emph{IN}_{flu}$ datasets.

Additionally, we generate another dataset, named $\emph{IN}_{test}$, using 10 different random seeds not included in $\emph{IN}_{train}$, including 10,000 images. All experiments are conducted on four NVIDIA GEFORCE RTX 2080 Ti GPU cards.

\subsection{Evaluation Metrics}

\subsubsection{Resistance to Steganalysis} The security performance of steganographic methods is usually evaluated using the detection error rate of steganalyzers. The detection error rate $P_{\mathrm{E}}$ is defined as: 
\begin{equation}
    P_{\mathrm{E}}= \frac{1}{2}\left(P_{\mathrm{FA}}+P_{\mathrm{MD}}\right),
\end{equation}
where $P_{\mathrm{FA}}$ and $P_{\mathrm{MD}}$ represent the false-alarm (FA) rate and the missed detection (MD) rate of steganalyzers, respectively.

\subsection{Experimental Results and Analysis}

\begin{figure*}[h]\centering
\includegraphics[width=\textwidth]{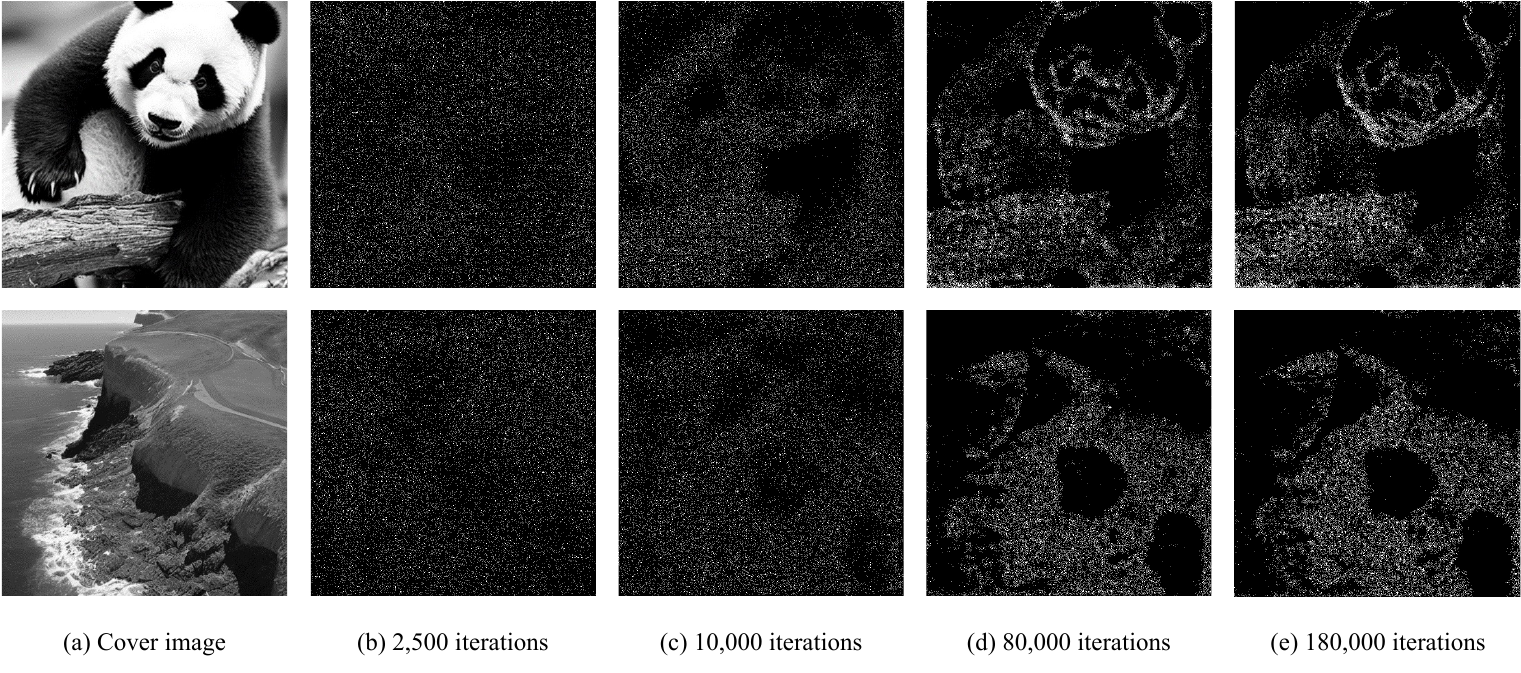}
\caption{Cover images (generated by Stable Diffusion) and modification maps with different number of iterations.}
\label{iter}
\end{figure*}

\begin{table*}[h]
\begin{center}
\caption{Detection error rate(\%) of SRM, CovNet, and LWENet under different $\lambda$. A higher rate indicates higher security.}
\renewcommand{\arraystretch}{1.5}
\setlength{\tabcolsep}{12pt}
\label{tab_lambda}
\begin{tabular}{ c | c | c | c | c| c| c }
\hline
\multirow{2}{*}{Steganalyzer}  &  \multirow{2}{*}{Parameter} &   \multirow{2}{*}{0.1 bpp}       & \multirow{2}{*}{0.2 bpp} & \multirow{2}{*}{0.3 bpp} & \multirow{2}{*}{0.4 bpp} & \multirow{2}{*}{Average}\\
&&&&&&\\
\hline

\multirow{5}{*}{SRM}& $\lambda$ = 1  & \textbf{46.74}& \textbf{44.09}&38.53 &33.99 &\textbf{40.84}\\

                       & $\lambda$ = 2& 46.49& 42.68&\textbf{39.43} &33.35 &40.49\\

                       &$\lambda$ = 4 &45.83	&42.08	&38.66	&33.80 & 40.09\\

                       & $\lambda$ = 8&46.09 &42.72 &38.92 &\textbf{34.42} &40.54\\

\hline
\multirow{5}{*}{CovNet}& $\lambda$ = 1  &\textbf{45.07} &\textbf{41.69} &35.31 & 29.40&\textbf{37.87}\\

                       & $\lambda$ = 2& 44.25&40.01 &\textbf{36.96} &29.66&37.72 \\

                       &$\lambda$ = 4 &43.41	&40.87	&34.20	&30.34 & 37.21\\

                       & $\lambda$ = 8&40.44 &39.46& 36.56 &\textbf{31.82}&37.07\\

\hline
\multirow{5}{*}{LWENet}&$\lambda$ = 1  & 43.35&\textbf{42.36} &36.49 &30.52 &\textbf{38.18}\\

                       & $\lambda$ = 2&42.58 &40.80 &\textbf{38.57} &30.45 &38.10\\

                       &$\lambda$ = 4 &\textbf{43.48}	&40.95	&34.48	&29.97 & 37.22\\

                       & $\lambda$ = 8& 40.17& 38.70&36.99 &\textbf{30.77} &36.66\\

\hline
\end{tabular}
\end{center}
\end{table*}

\subsubsection{The optimal value of the parameter $\tau$}

In the design of GIFDL, the adversarial loss $l_a$ of the Generator consists of two parts that constrain the generator in two ways: (1) the stego image should be indistinguishable from the cover image, and (2) the stego image should be indistinguishable from the randomly chosen fluctuation image. To balance the impact of two constraints on steganographic security, we introduce a parameter $\lambda$ in section~\ref{loss} to regulate the contribution of $E_2(F_{ran}, S)$ to the total generator loss $l_a$. To determine the optimal value for $\lambda$, we conduct the following experiment: 

We set the value of $\lambda$ to 1, 2, 4, and 8, respectively. After training GIFDL, we use 10,000 grayscale images from the $\emph{IN}_{test}$ dataset as the cover images for testing the security of GIFDL. Specifically, we generate the corresponding steganographic cost through GIFDL and embed the secret message via STC, with the payload set to 0.4 bpp. We obtain 10,000 stego images via the above operations. We then divide the 10,000 cover-stego image pairs into three subsets of sizes 4,000, 1,000, and 5,000, which are used as the training, validation, and test datasets, respectively.

Finally, we train and test three steganalyzers (SRM~\cite{srm}, CovNet~\cite{covnet}, and LWENet~\cite{lwenet}) using the above datasets, and record the detection error rate ($P_E$) on the test dataset. The experimental results, shown in Table~\ref{tab_lambda}, indicate that when $\lambda=1$, GIFDL exhibits the best resistance against the three steganalyzers, achieving optimal security performance. Therefore, in all subsequent experiments, we set the parameter $\lambda$ to 1 for GIFDL. Additionally, we observe that a larger $\lambda$ results in higher security as the payload increases. This phenomenon can be explained by the fact that a larger $\lambda$ gives greater weight to the role of fluctuation images, and the fluctuation distribution can provide more modification regions for steganographic modifications, which are also relatively secure, i.e., a larger $\lambda$ offers more optional regions for steganography. When the payload increases, the security is increased compared with the lower $\lambda$ because there are more ``secure” locations to modify.

As shown in Fig.~\ref{iter}, as the number of training iterations of GIFDL increases, the steganographic modifications shift from random modifications to concentrate on regions with complex image textures. This indicates that the training is effective and that the performance of the generator gradually improves.

\subsubsection{Security Performance}

\begin{figure*}[t]\centering
\includegraphics[width=\textwidth]{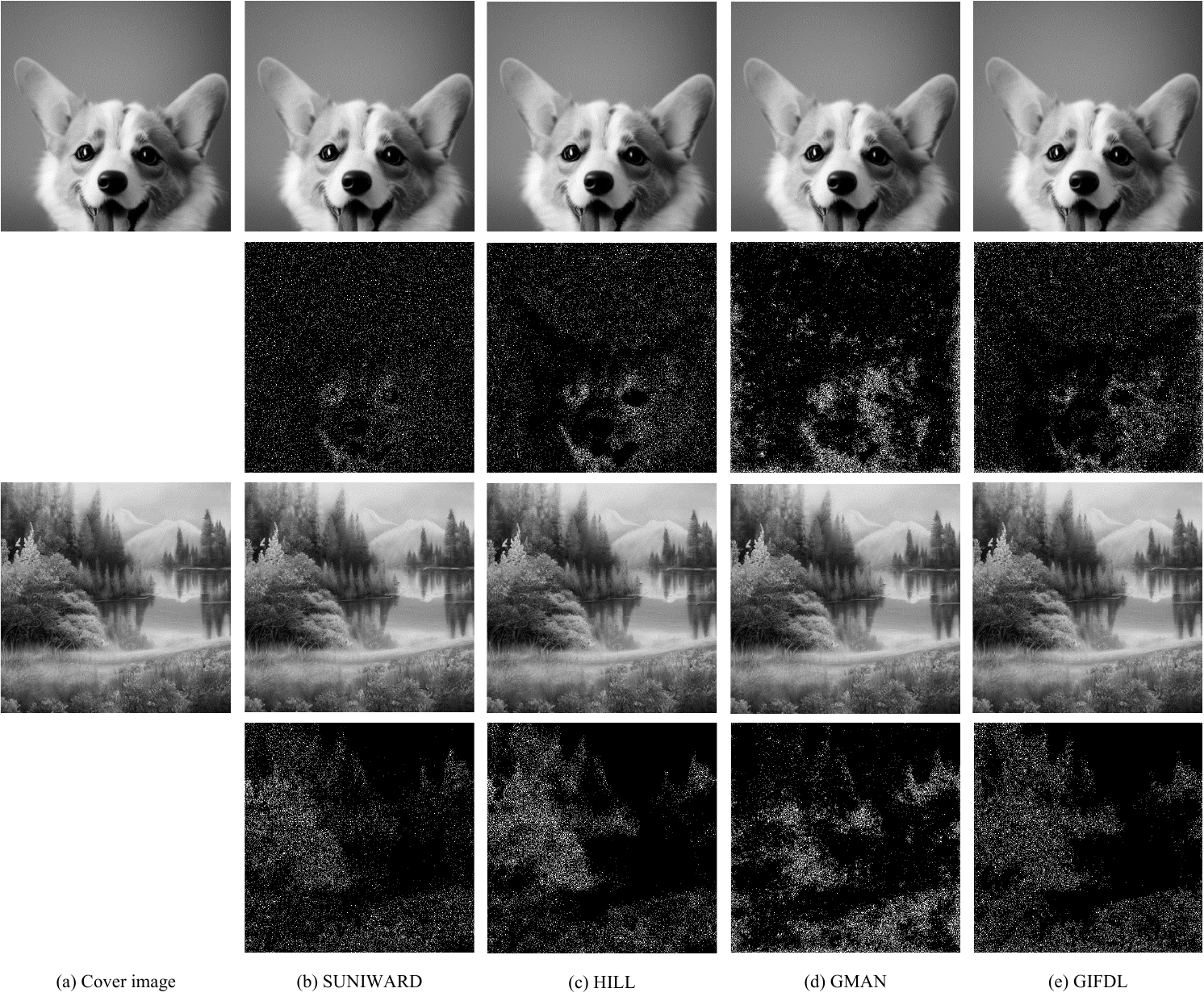}
\caption{Cover images, stego images (first row, third row), and corresponding modification maps (second row, fourth row).}
\label{compare}
\end{figure*}

\begin{table*}[t]
\begin{center}
\caption{Detection error rate(\%) of SRM, CovNet, and LWENet under different payload. A higher rate indicates higher security.}
\renewcommand{\arraystretch}{1.5}
\setlength{\tabcolsep}{12pt}
\label{tab_compare}
\begin{tabular}{ c | c | c | c | c| c| c }
\hline
\multirow{2}{*}{Steganalyzer}  &  \multirow{2}{*}{Method} &   \multirow{2}{*}{0.1 bpp}       & \multirow{2}{*}{0.2 bpp} & \multirow{2}{*}{0.3 bpp} & \multirow{2}{*}{0.4 bpp} & \multirow{2}{*}{Average}\\
&&&&&&\\
\hline

\multirow{6}{*}{SRM}& S-UNIWARD  & 42.20& 38.28& 32.66& 27.29 &35.11\\

                       & HILL& 43.70&37.15 &33.24 &27.72&35.45 \\

                       &GACL &	44.16&	38.62&	35.57& 33.68&38.01 \\
                       
                       &GMAN &45.43	&41.57	&34.16	&30.78 & 37.99\\

                       & GIFDL& \textbf{46.74}& \textbf{44.09}&\textbf{38.53} &\textbf{33.99} &\textbf{40.84}\\

\hline
\multirow{6}{*}{CovNet}& S-UNIWARD  &38.65 &33.06 &25.56 &18.87&29.04 \\

                       & HILL& 39.09& 32.70&27.23 &20.73&29.94 \\

                       &GACL &39.95	&34.10	&33.89	&26.95 &33.72 \\
                       
                       &GMAN &41.59	&33.64	&30.69	&27.10 & 33.26\\

                       & GIFDL&\textbf{45.07} &\textbf{41.69} &\textbf{35.31} & \textbf{29.40}&\textbf{37.87}\\

\hline
\multirow{6}{*}{LWENet}& S-UNIWARD  &36.20 &30.24 &27.15 &21.25&28.71 \\

                       & HILL&39.56 &31.31 &29.18 &23.12&30.79 \\

                       &GACL &40.57	&36.94	&28.64	&22.44 &32.15 \\
                       
                       &GMAN &42.43	&39.64	&33.66	&27.25 & 35.75\\

                       & GIFDL& \textbf{43.35}&\textbf{42.36} &\textbf{36.49} &\textbf{30.52} &\textbf{38.18}\\

\hline
\end{tabular}
\end{center}
\end{table*}

In this section, we compare the security performance of GIFDL with that of other steganographic cost learning methods such as SUNIWARD~\cite{ref2}, HILL~\cite{ref3}, GMAN~\cite{GMAN}, and GACL~\cite{wang2024gan}. Among them, the distortion definitions of SUNIWARD and HILL are heuristic and do not require pretraining. First, we train GMAN and GACL using the $\emph{IN}_{train}$ dataset, and the proposed GIFDL using the $\emph{IN}_{train}$ and $\emph{IN}_{flu}$ datasets. Next, we use the $\emph{IN}_{test}$ dataset as the cover images, and use SUNIWARD, HILL, GMAN, GACL, and GIFDL, respectively, with payloads of 0.1 bpp, 0.2 bpp, 0.3 bpp, 0.4 bpp respectively to obtain the corresponding stego images. The stego images and modification maps of the cover images with the four steganography methods are shown in Fig.~\ref{compare}. Unlike other steganography methods, the steganographic modifications of GIFDL are not only concentrated in the complex regions of the image texture but also distributed in the regions of fluctuation, which is very similar to Fig.~\ref{pmoti}(b). We use the above cover-stego pairs to train three steganalyzers, SRM, CovNet, and LWENet, respectively. where the cover-stego pairs are divided into three datasets of sizes 4000, 1000, and 5000, which are used for steganalyzer training, validation, and testing, respectively. The detection error rates ($P_E$) of the steganalyzers are shown in Table.~\ref{tab_compare}, where it can be observed that all four steganographic methods are more easily detected by the steganalysis as the payload increases. Moreover, $P_E$ for GIFDL is significantly lower than that of the comparison methods, and the average $P_E$ is 3.30\% higher than that of the suboptimal GMAN. Therefore, GIFDL has the highest security in terms of resistance to steganalysis.

\subsubsection{Security performance on real datasets}

\begin{figure}[t]\centering
\includegraphics[width=\columnwidth]{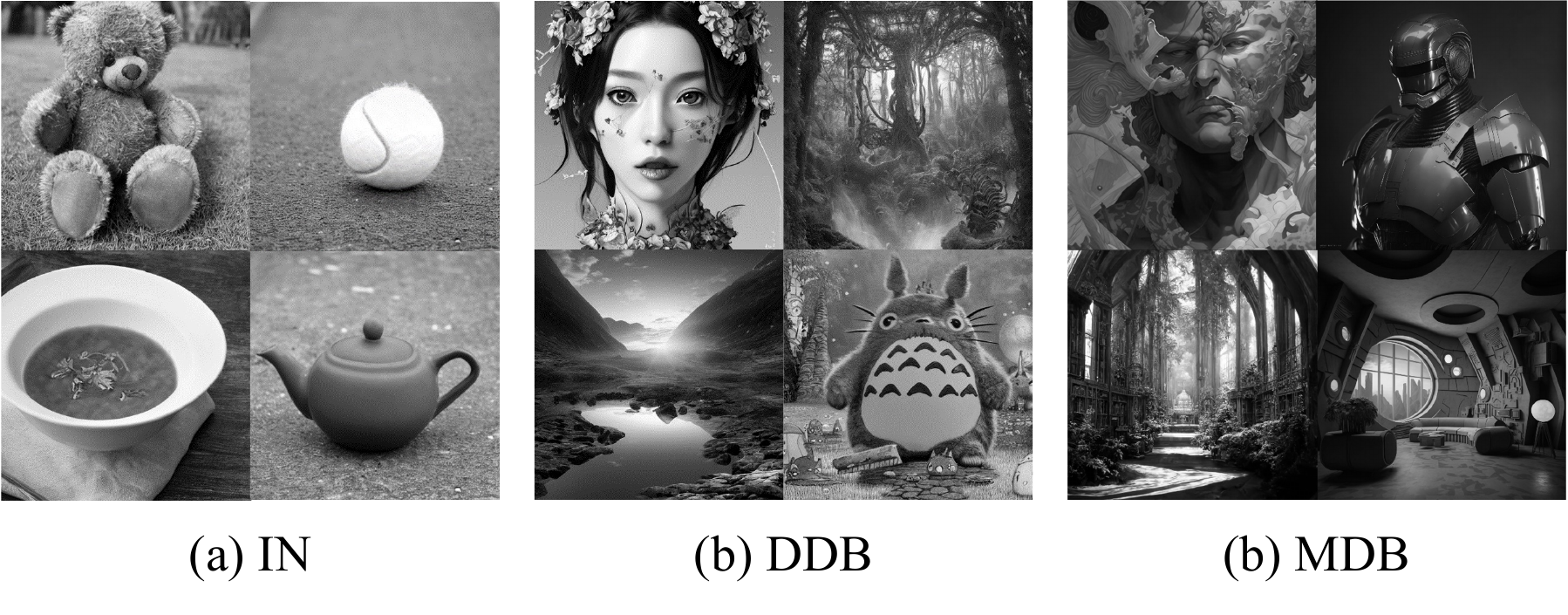}
\caption{Differences between datasets $\emph{IN}$, $\emph{DDB}$ and $\emph{MDB}$. (a) The prompts used in $\emph{IN}$ come from class labels of ImageNet, and the images in $\emph{IN}$ are usually single objects with simpler textures. (b) The prompts used in $\emph{DDB}$ and \emph{MDB} are from real users, and the images in $\emph{DDB}$ are more artistic and have more complex textures.}
\label{ddb}
\end{figure}
In this section, we explore the generalization performance of GIFDL on different datasets. We use the dataset $\emph{IN}_{train}$ to train GIFDL and the comparison methods. Then we test them on the dataset DiffusionDB~\cite{wang2022diffusiondb} and JourneyDB~\cite{sun2023journeydb}. DiffusionDB is the first large-scale text-to-image prompt dataset. It contains 14 million images generated by Stable Diffusion using prompts and hyperparameters specified by real users. We randomly select 10,000 images of size 512$\times$ 512 from the DiffusionDB dataset to form our test dataset, which is named $\emph{DDB}$. Meanwhile, We obtain 10,000 Midjourney-generated images from the JourneyDB dataset. These images have a resolution of 1024×1024 and are then resized to 512×512 using the $imresize$ function with the bicubic interpolation method and the default parameter settings in MATLAB. We name the above dataset the Midjourney Database (MDB). Fig.~\ref{ddb} shows the main differences between the training set $\emph{IN}$ and the test set $\emph{DDB}$, $\emph{MDB}$. The images in $\emph{DDB}$ and $\emph{MDB}$ usually along with longer and more complex prompts, and the models used are usually stable diffusion v1.5, v2.0, and Midjourney. We use the $\emph{DDB}$ and $\emph{MDB}$ datasets as cover images and use GIFDL and comparison methods to obtain the corresponding stego images, respectively. Cover-stego pairs are partitioned into three datasets of sizes 4000, 1000, and 5000, which are used for training, validation, and testing of the steganalyzers, respectively. The experimental results are shown in Table~\ref{tab_ddb}, where it can be observed that GIFDL maintains the resistance to steganalysis performance on the real dataset and possesses good generalizability to the stable diffusion family of models. Meanwhile, GIFDL maintains a certain generalization performance on the image generation model Midjourney. Although the performance of GIFDL shows some degradation on Midjourney, it is still significantly better than the comparison methods.

\begin{table}[h]
\begin{center}
\renewcommand{\arraystretch}{1.5}
\setlength{\tabcolsep}{10pt}
\caption{Detection error rates(\%) of SRM, CovNet, and LWENet on different datasets. A higher rate indicates higher security.}
\label{tab_ddb}
\begin{tabular}{ c |c| c|  c| c }
\hline
 Dataset& Method &SRM & CovNet& LWENet\\
\hline
\multirow{5}{*}{$\emph{IN}$}&SUNIWARD& 27.29 & 18.87 & 21.25 \\

&HILL & 27.72& 20.73 & 23.12 \\

&GACL &33.68 &26.95  &  22.44\\

&GMAN &30.78 & 27.10 & 27.25 \\

&GIFDL  &\textbf{33.99} & \textbf{29.40} & \textbf{30.52} \\
\hline
\multirow{5}{*}{$\emph{DDB}$}&SUNIWARD&27.86 & 23.77 & 23.46 \\

&HILL &31.10 & 27.11 &  26.56\\

&GACL &33.44 &28.58  &  24.06\\

&GMAN &30.31 & 27.51 &  28.97\\

&GIFDL  &\textbf{35.85} & \textbf{31.60} & \textbf{33.24} \\
\hline
\multirow{5}{*}{$\emph{MDB}$}&SUNIWARD&23.10 &19.45  &16.21  \\

&HILL &\textbf{26.56} &22.40  &16.63  \\

&GACL &24.19 &18.25  &15.27 \\

&GMAN &23.79 &19.31  &17.99  \\

&GIFDL  &26.34 & \textbf{25.99} & \textbf{24.99} \\
\hline
\end{tabular}
\end{center}
\end{table}

\subsubsection{Ablation study}
\label{aban}

\begin{table}[h]
\begin{center}
\caption{Several variants of GIFDL.}
\label{tab_ab}
\begin{tabular}{ c  c  c c }
\toprule
Variant &  Fluctuation image  & Threshold & Discriminator strategy \\
\midrule
Variant \#1& $\times$  & $\times$ & $\times$ \\
\midrule
Variant \#2 &  $\checkmark$  & $\times$ & $\times$ \\
\midrule
Variant \#3 & $\checkmark$ & $\checkmark$ &  $\times$\\
\midrule
Variant \#4 & $\checkmark$ & $\checkmark$ & $\checkmark$ \\
\bottomrule 
\end{tabular}
\end{center}
\end{table}

In this section, we explore the role of each component of GIFDL. For this purpose, we design several variants of GIFDL in Table~\ref{tab_ab}. Among them, Variant \#1 refers to GMAN, ``Fluctuation image" denotes the use of fluctuation images during training, and ``Threshold" denotes the use of threshold $\tau$ to filter the fluctuation images. ``Discriminator strategy" indicates that the assigned discriminator update strategy is used, i.e., two different steganalyzers are used as two discriminators to discriminate cover-stego pairs and fluctuation-stego pairs, respectively. If this item is unchecked, it means that two identical steganalyzers are used as two discriminators to discriminate cover-stego pairs and fluctuation-stego pairs, respectively. We use the dataset $\emph{IN}_{train}$ to train the above variants and test the performance of these variants against steganalysis on the dataset $\emph{IN}_{test}$, and the experimental results are shown in Table~\ref{tab_abr}. 

Based on the experimental results of Variant \#1 and Variant \#2, using fluctuation images without a threshold to constrain their differences from cover images during training does not significantly improve the security of steganographic images. This can be explained from two perspectives. First, the majority of fluctuation images and cover images are visually indistinguishable, so their differences are concentrated in regions with complex textures. These images are involved in the training process, which helps to improve steganographic security. Second, without a threshold restriction, some fluctuation images with substantial differences from the cover image (e.g., Fig.~\ref{pgm}) are included in the training. When these highly different images are involved in training, GIFDL evaluates the content change regions as suitable areas for steganographic modification. However, in reality, these regions are very smooth and not suitable for steganographic modification. These highly different images neutralize the gains brought by learning distortion from similar fluctuation images, which is why the experimental results of Variant \#1 and Variant \#2 are similar. When we introduce a threshold into the framework (i.e., Variant \#3), these differentiated images are excluded, and the training set only includes fluctuation images that we consider ideal. Naturally, the performance is improved as a result. The experimental results for Variant \#3 and Variant \#4 demonstrate that the steganographic security can be effectively improved by using the assigned discriminator update strategy, which suggests that it is reasonable to use two discriminators that differ in performance. In fact, by observing the change in the loss of the discriminator in Variant \#3, it can be found that the loss of the discriminator used to discriminate the fluctuation-stego pairs always reduces to 0 very quickly, which results in the generator not being able to learn effective knowledge during the adversarial process, i.e., the gradient disappears. In conclusion, the use of ``Fluctuation image" can provide additional information for steganography, which can improve the steganographic security under the ``Threshold" constraint. Moreover, the allocation-based discriminator update strategy can balance the contributions of cover images and fluctuation images to further improve steganographic security.

\begin{table}[h]
\begin{center}
\caption{Detection error rates(\%) of SRM, CovNet, and LWENet under different variants. A higher rate indicates higher security.}
\label{tab_abr}
\begin{tabular}{ c  c  c c }
\toprule
Variant &  SRM  & CovNet & LWENet \\
\midrule
Variant \#1&  30.78 & 27.10 & 27.25\\
\midrule
Variant \#2 &  31.81  & 26.80 & 27.45 \\
\midrule
Variant \#3 & 32.56 & 29.03 & 29.71 \\
\midrule
Variant \#4 &\textbf{33.99} & \textbf{29.40} & \textbf{30.52}  \\
\bottomrule 
\end{tabular}
\end{center}
\end{table}

\subsubsection{An alternative discriminator training strategy}
\label{d2}
In section~\ref{c2-b}, our design of the discriminator is such that two different discriminators are applied to different discrimination tasks. That is, Yed-Net is used to discriminate between cover and stego, and XuNet is used to discriminate between flu and stego. It has been shown in section~\ref{aban} that the use of two discriminators with different performances better balances the contributions of the cover and the fluctuation images, compared with the use of the same discriminator. However, there is naturally a more mundane design where two discriminators are applied to the same discrimination task. This is done by Yed-Net and XuNet, which are used to discriminate not only cover and stego but also flu and stego. Thus, we redefine the discriminator loss:
\begin{equation}
\begin{split}
   l_D = max\{E_i(F_{ran},S)+\lambda'  E_i(C,S), i=1,2\},
   \end{split}
\end{equation}
\begin{equation}
\begin{split}
   l_a = min\{E_i(F_{ran},S)+\lambda'  E_i(C,S), i=1,2\},
   \end{split}
\end{equation}where we use a different training strategy: the two discriminators have the same classification task, i.e., discriminating both cover-stego pairs and fluctuation-stego pairs, using the parameter $\lambda'$ to weigh the contributions of both. In each epoch, we update the parameters of the discriminator with the higher loss, leaving the other discriminator unchanged. At the same time, the adversarial loss $l_a$ of the steganography generator is the smaller loss. We call this experimental setup GIFDL*.

We use the GIFDL and GIFDL* models with $\lambda'$ values of 1, 2, 4, and 8 to generate 10,000 stego images respectively, resulting in five stego datasets. For each stego dataset, we divide the 10,000 stego images into training, validation, and test datasets, with sizes of 4000, 1000, and 5000, respectively. Subsequently, we use these datasets to train three state-of-the-art steganalysis networks, SRM, CovNet, and LWENet. The experimental results are presented in Table~\ref{tab2}. This mundane design method does not achieve better security performance, which can be explained by the fact that the simultaneous execution of two tasks by each discriminator leads to a degradation of its performance under both tasks. In contrast, our proposed method of assigning different discriminators to each of the two tasks is able to maximize the performance of the discriminators, and thus the performance of the generator, while taking into account the differences in the cover and the fluctuation images.

\begin{table}[h]
\begin{center}
\caption{Detection error rates(\%) of SRM, CovNet, and LWENet under different strategies. A higher rate indicates higher security.}
\label{tab2}
\begin{tabular}{ c  c  c c }
\toprule
Method &  SRM & CovNet& LWENet\\
\midrule
GIFDL&  \textbf{33.99}  & \textbf{29.40} & \textbf{30.52} \\
\midrule
GIFDL* with $\lambda' =1$ & 31.96 & 27.55 & 26.66 \\
\midrule
GIFDL* with $\lambda' =2$ & 32.10 & 26.21 &  27.24\\
\midrule
GIFDL* with $\lambda' =4$ & 32.85 & 28.60 &  28.93\\
\midrule
GIFDL* with $\lambda' =8$ & 31.91 & 27.95 &  27.31 \\
\bottomrule 
\end{tabular}
\end{center}
\end{table}

\subsubsection{Combined with Volatility Cost}

In this section, we combine GIFDL and GMAN with the volatility cost (VC) proposed in~\cite{zjs} as a way to explore the effectiveness of GIFDL in capturing the fluctuations of generated images via deep learning networks. To this end, we generate steganographic costs on the dataset $\emph{IN}_{test}$ via GMAN and GIFDL, respectively, and compute the volatility costs of the dataset $\emph{IN}_{test}$ in the manner presented in~\cite{zjs}. The original steganographic cost is denoted as $\rho^o$, the volatility cost as $\rho^v$, and the steganographic cost after combining is denoted as $\rho^c$. The above process can be expressed as follows:

\begin{equation}
\begin{array}{cl} 
\rho^c(+1)=\beta \cdot \rho^v(+1) + (1-\beta) \cdot (\alpha \cdot \rho^o(+1)), \vspace{1.5ex} \\
\rho^c(-1)=\beta \cdot \rho^v(-1) + (1-\beta) \cdot (\alpha \cdot \rho^o(-1)),
\end{array}
\label{combine}
\end{equation}
where $\rho^c(+1)$ and $\rho^c(-1)$ represent the costs associated with +1 modification and -1 modification, $\beta$ is the hyperparameter that determines the proportion of volatility cost, and $\alpha$ is the scaling factor that equalizes the mean of $\rho^o$ with the volatility cost $\rho^v$. $\alpha$ is calculated as follows:
\begin{equation}
\alpha =   \frac{{\textstyle \sum_{i,j}\left ( [\rho_{ij}^v\ne \mathrm{wetcost}] \cdot \rho_{ij}^v  \right )}  /{\textstyle \sum_{i,j}[\rho_{ij}^v\ne\mathrm{wetcost}]}}{{\textstyle \sum_{i,j} \left (  [\rho_{ij}^o\ne\mathrm{wetcost}] \cdot \rho_{ij}^o\right ) }/{\textstyle \sum_{i,j}[\rho_{ij}^o\ne\mathrm{wetcost}]  }},
\end{equation}
where $\alpha$ represents the ratio between the average value of the volatility cost and the original cost, the Iverson bracket $[Q]$ is defined to be 1 if the logical expression $Q$ is true and 0 otherwise, and ``wetcost'' represents the cost tending towards infinity, which is excluded to avoid substantial impacts on the mean. It has been experimentally demonstrated in~\cite{zjs} that optimal performance is achieved with $\beta = 0.15$, so we take $\beta = 0.15$.

We combine the volatility cost with the steganographic cost of GMAN, named GMAN+VC, and similarly, combine the volatility cost with the steganographic cost of GIFDL, named GIFDL+VC. We utilize the above costs to embed the secret message via STC~\cite{stc} to obtain the corresponding stego image. The detection error rates of the three steganalyzers are shown in Table~\ref{tab_vc}. Compared with GMAN, the steganalysis error rate of GMAN+VC is increased by 8.77\% on average. Also, GMAN+VC improves by 5.85\% on average compared to GIFDL. In addition, GIFDL+VC improves 9.16\% on average compared to GIFDL.

In VC proposed by Zhang et al.~\cite{zjs}, the pixel differences between fluctuation images are modeled as Gaussian distributions. However, there is a difference between the actual distribution and the standard Gaussian distribution. In GIFDL, we directly learn the pixel distribution of fluctuation images from training data through a deep learning network, rather than simply assuming it to be a Gaussian distribution. Due to the differences in method design, new knowledge can possibly be learned, and experimental results have also verified that GIFDL and VC can be complementary.

\begin{table}[h]
\begin{center}
\caption{Detection error rates(\%) of SRM, CovNet, and LWENet under different combinations. A higher rate indicates higher security.}
\label{tab_vc}
\begin{tabular}{ c  c  c c  c}
\toprule
Method &  SRM & CovNet& LWENet & Average\\
\midrule
GMAN& 30.78 & 27.10 & 27.25  &28.38\\
\midrule
GMAN + VC & 36.09 & 38.86 &  36.51 &37.15\\
\midrule
GIFDL & 33.99 & 29.40 & 30.52 & 31.30\\
\midrule
GIFDL + VC& \textbf{41.27} & \textbf{40.64} &  \textbf{39.46}&  \textbf{40.46}\\
\bottomrule 
\end{tabular}
\end{center}
\end{table}

\section{Conclusion}
\label{c4}

To further enhance steganographic security, we propose a steganographic distortion learning method based on the fluctuations of generated images. Specifically, we observe that generated images exhibit a fluctuation distribution, and we use this as side information to guide the generator in learning steganographic distortion. To avoid the problem of gradient vanishing during training and to fully utilize the performance of the discriminators, we introduce a training strategy called ``assignment", which assigns different tasks to the two discriminators and balances their performance by updating the parameters alternately. Experimental results indicate that compared with GMAN, GIFDL can further improve steganographic security and has some generalizability, maintaining performance on new datasets.

In our experiments, we observed that the differences between fluctuation images and cover images are not entirely concentrated in areas with complex textures; differences often exist in the background regions as well. This is due to the randomness introduced by the noise process in DDPMs. Future research will focus on exploring whether these random differences could potentially reduce steganographic security. Additionally, we will investigate how to better utilize the fluctuation characteristics of generated images, considering a more refined estimation of their distribution. Overall, we propose a new method to enhance steganographic security using generated images and explore the direction of steganography via black-box generative models.

\bibliographystyle{IEEEtran}

\end{document}